\documentclass[prl,twocolumn,superscriptaddress,nobibnotes,letterpaper]{revtex4-1}

\usepackage[pdftex]{graphicx}
\usepackage[pdftex]{epsfig}
\usepackage{amsmath}
\usepackage{amssymb}
\usepackage{amsfonts}
\usepackage{color}
\usepackage{wrapfig}
\usepackage{eucal}
\usepackage{hhline}
\usepackage{threeparttable}
\usepackage{supertabular}
\usepackage{multirow}
\usepackage{tabularx}

\usepackage[colorlinks=true, allcolors=blue]{hyperref}

\def\be{\begin{equation}}
\def\ee{\end{equation}}
\def\bea{\begin{eqnarray}}
\def\eea{\end{eqnarray}}

\newcolumntype{Y}{>{\centering\arraybackslash}X}

\begin{document}

\pagenumbering{arabic}

\title{Optomechanically induced optical trapping system based on photonic crystal cavities}

\author{Manuel Monterrosas-Romero}
\affiliation{Institute for Functional Matter and Quantum Technologies, Universität Stuttgart, 70569 Stuttgart, Germany}
\affiliation{Center for Integrated Quantum Science and Technology (IQST), University of Stuttgart, 70569 Stuttgart, Germany}

\author{Seyed K. Alavi}
\affiliation{Institute for Functional Matter and Quantum Technologies, Universität Stuttgart, 70569 Stuttgart, Germany}
\affiliation{Center for Integrated Quantum Science and Technology (IQST), University of Stuttgart, 70569 Stuttgart, Germany}

\author{Ester M. Koistinen}
\affiliation{Institute for Functional Matter and Quantum Technologies, Universität Stuttgart, 70569 Stuttgart, Germany}

\author{Sungkun Hong}
\email{sungkun.hong@fmq.uni-stuttgart.de}
\affiliation{Institute for Functional Matter and Quantum Technologies, Universität Stuttgart, 70569 Stuttgart, Germany}
\affiliation{Center for Integrated Quantum Science and Technology (IQST), University of Stuttgart, 70569 Stuttgart, Germany}

% \address{\authormark{1}Peer Review, Publications Department, Optica Publishing Group, 2010 Massachusetts Avenue NW, Washington, DC 20036, USA\\
% \authormark{2}Publications Department, Optica Publishing Group, 2010 Massachusetts Avenue NW, Washington, DC 20036, USA\\
% \authormark{3}Currently with the Department of Electronic Journals, Optica Publishing Group, 2010 Massachusetts Avenue NW, Washington, DC 20036, USA}

% use {asbstract*} to suppress the copyright line. Copyright information will be added in production

\begin{abstract} 
Optical trapping has proven to be a valuable experimental technique for precisely controlling small dielectric objects. However, due to their very nature, conventional optical traps are diffraction limited and require high intensities to confine the dielectric objects. In this work, we propose a novel optical trap based on dielectric photonic crystal nanobeam cavities, which overcomes the limitations of conventional optical traps by significant factors. This is achieved by exploiting an optomechanically induced backaction mechanism between a dielectric nanoparticle and the cavities. We perform numerical simulations to show that our trap can fully levitate a submicron-scale dielectric particle with a trap width as narrow as 56 nm. It allows for achieving a high trap stiffness, therefore, a high Q-frequency product for the particle’s motion while reducing the optical absorption by a factor of 43 compared to the cases for conventional optical tweezers. Moreover, we show that multiple laser tones can be used further to create a complex, dynamic potential landscape with feature sizes well below the diffraction limit. The presented optical trapping system offers new opportunities for precision sensing and fundamental quantum experiments based on levitated particles.
\end{abstract}

\maketitle

%%%%%%%%%%%%%%%%%%%%%%%%%%  body  %%%%%%%%%%%%%%%%%%%%%%%%%%
Optical trapping is a versatile tool in modern science. Since its birth \cite{Ashkin.1970}, optical trapping has been used in various groundbreaking experiments across disciplines, ranging from trapping and cooling of atoms \cite{Phillips.1998} to the manipulation of individual living cells \cite{Ashkin.1987}. Recently, optical trapping has also found its utility in quantum optomechanics \cite{Chang.2010, RomeroIsart.2010}. In high vacuum, a dielectric nanoparticle trapped in an optical tweezer becomes an excellent mechanical oscillator with an ultrahigh quality factor Q. It has allowed the observation and control of the particle’s motion at the quantum limit \cite{Delic.2020, Magrini.2021, Tebbenjohanns.2021}, paving the way for new sensing technologies \cite{Moore.2014, Ranjit.2016, Rider.2016} and probing quantum physics in new mass and length scales \cite{RomeroIsart.2011, Yin.2013, Scala.2013}.

\begin{figure*}[t]
	\begin{center}
		\includegraphics[width=1.4\columnwidth]{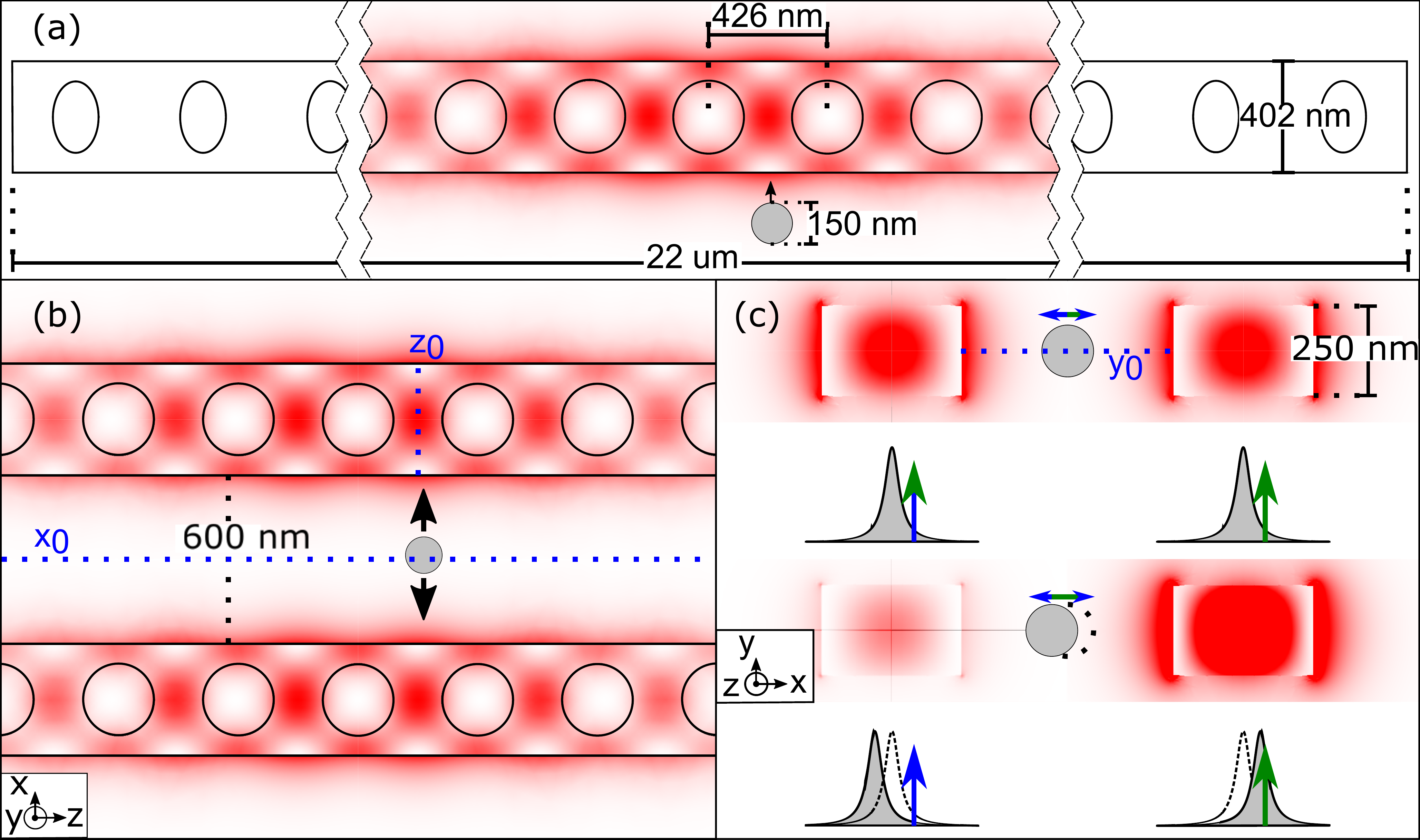}
		\caption{Working principle of the PCNC-based SIBA trap \textbf{a.} Simulated electric field intensity of a silicon photonic crystal nanobeam cavity (PCNC) used in our study. The PCNC is designed to be 22 um in length and 402 nm in width. At each side of the PCNC, an array of holes is created to form a Bragg mirror. These hole arrays are extended to the center of the PCNC with the hole size and spacing properly tapered to form a localized cavity mode. When a 150 nm silica nanosphere (a light gray circle) is brought near a single PCNC, the particle is attracted to the PCNC due to the optical gradient force by the cavity’s evanescent field. \textbf{b.} A layout of a PCNC-based self-induced backaction (SIBA) trap. Two PCNCs are placed in parallel with a distance of 600 nm. The inset in the bottom left corner shows the coordinate system used throughout the article. The $x$ and $z$ coordinates of its origin, ($x_0$, $y_0$, $z_0$), are also depicted; the plane of $x = x_0$ lies precisely in the middle between the two PCNCs, and the $z = z_0$ plane lies in between the cavity hole at the center and the neighboring hole, where the maximum of the cavity’s evanescent field is located. The $y = y_0$ plane cuts through the middle of the PCNCs in the $y$ direction (shown in (c)). \textbf{c.} Schematic illustration of restoring optical forces arising from SIBA effect. A 150 nm silica nanosphere is initially positioned at the origin ($x_0$, $y_0$, $z_0$). When the two PCNCs are pumped with equal laser power and detuning, the optical pulling forces on the particle are balanced. When the particle is displaced to one of the PCNCs (e.g., to the left), the SIBA effect shifts the cavities’ resonance frequencies such that the cavity field intensity on the left (right) is substantially reduced (increased), creating an imbalance in the magnitude of the two optical pulling forces. It results in a net force pushing the particle back toward the origin, keeping the particle in between the two PCNCs.}
		\label{fig:1}
	\end{center} 
\end{figure*}

However, there also exist outstanding challenges to further advancing tweezer-based quantum optomechanics. Realizing a high Q-frequency product of the mechanical oscillator \cite{Tsaturyan.2017, Ghadimi.2018} is an essential prerequisite for performing precision and quantum-coherent experiments. With standard optical tweezers, it is achieved by increasing the tweezer beam’s intensity, thus, the stiffness of the trap in which the particle oscillates. However, the intense laser field often also causes excessive absorption heating of the particle, in particular in high vacuum, and can result in the instability or even loss of the particle \cite{Frangeskou.2018}. Even if the particle survives the heating-induced instability, the increased blackbody radiation could severely limit the system’s coherence, thus precluding quantum experiments that require exceptionally long coherence times \cite{RomeroIsart.2011, Yin.2013, Scala.2013}. The trap stiffness can alternatively be enhanced by reducing the width of the trap. However, this possibility is also limited for a conventional optical trap, as the length scale over which the trap can vary is bound by the diffraction limit.

In the meantime, researchers have studied a different trapping mechanism that could potentially circumvent the abovementioned limitations \cite{Barth.2006, Juan.2009, Descharmes.2013}. It utilizes optical nanocavities as a means to provide localized optical trapping fields. When external lasers pump the cavities, the optical gradient forces generated by the cavity fields can attract and trap particles nearby. The key difference to conventional optical tweezers is that the particles also strongly affect the trapping fields by shifting the cavities’ resonance frequencies. This optomechanically-induced dynamic effect, also termed self-induced backaction (SIBA), results in an optical trap that is qualitatively different from standard optical tweezers. The SIBA effect has been first observed experimentally with a plasmonic nanocavity \cite{Juan.2009} and later with a dielectric photonic crystal cavity \cite{Descharmes.2013}. Recently, SIBA-based optical trapping was theoretically investigated in the context of optomechanics \cite{Neumeier.2015}. In this study, Neumeier et al. \cite{Neumeier.2015} have considered a simple Fabry-Perot cavity model and shown that the SIBA effect can produce optical traps with nontrivial shapes and sub-diffraction features if the cavity supports strong optomechanical interaction and sharp optical resonance. The next step is to devise a concrete cavity system that satisfies the required conditons for realizing low-intensity and high-stiffness optical levitation. 

\begin{figure*}[t]
\centering\includegraphics[width=1.4\columnwidth]{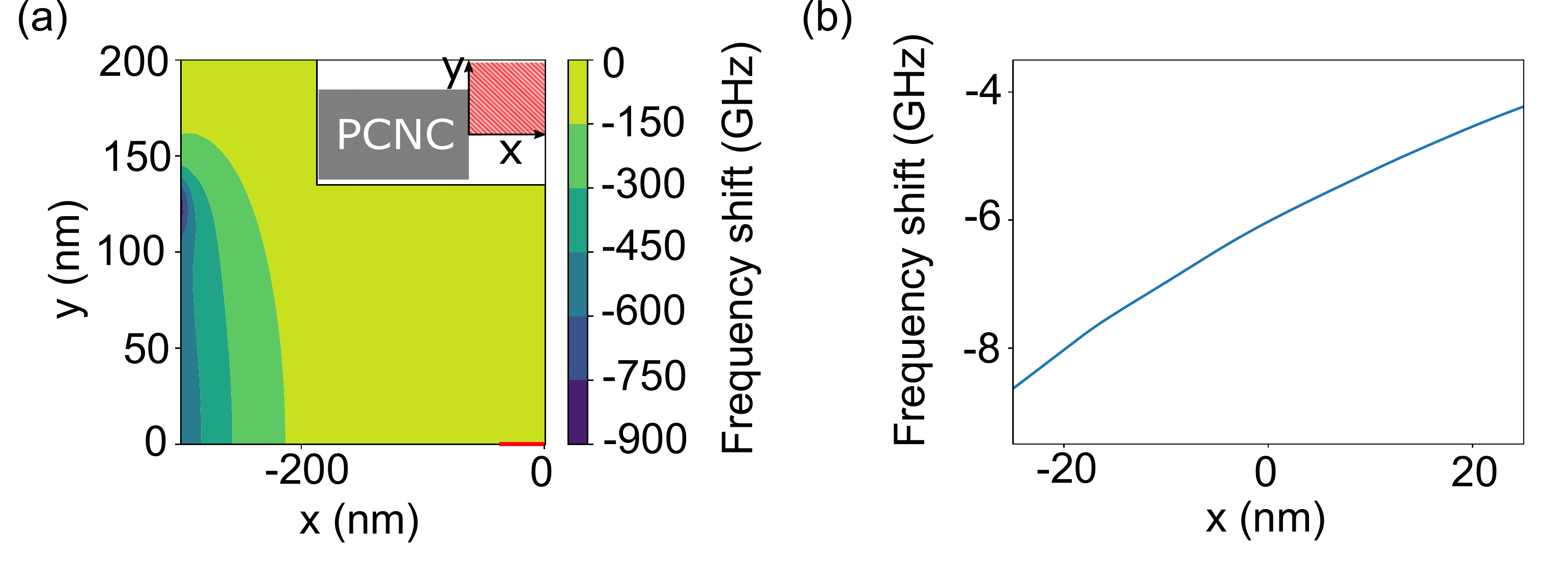}
\caption{Optomechanical coupling between the PCNC and the particle \textbf{a.} Shown is the map of the PCNC’s resonance frequency shift caused by the nanoparticle on a plane at $z=z_0$. The inset indicates the relative location between the mapped area and the PCNC. \textbf{b.} The line cut of the frequency shift zoomed in around the origin. The location and the span of the line cut are depicted as a red line near the bottom-right corner in (a). The particle’s displacement of 10 nm around $x = 0$ is enough to shift the cavity’s frequency by around 1 GHz.}
\label{fig:2}
\end{figure*}

In this article, we present a nanophotonic cavity system that can practically realize a high-stiffness SIBA trap for a dielectric nanoparticle. Our trap is based on two photonic crystal nanobeam cavities (PCNC) that can be conveniently made using conventional nanofabrication technologies. The PCNC’s high Q factor and the strong optomechanical response can result in enhanced SIBA effects that substantially modify the optical gradient force by the cavity field. We show that two PCNCs arranged in parallel, when pumped with appropriate laser fields, can form an optical trap with a width close to 50 nm. It dramatically reduces the optical field intensity required for a desired trap stiffness by a factor larger than 40. In contrast to previously demonstrated SIBA traps \cite{Juan.2009, Descharmes.2013}, the particle in our trap is fully levitated without any physical contact with the PCNCs. The particle can thus attain excellent mechanical coherence and stability in high vacuum. We also demonstrate the capability of our PCNC-based trap to create a more complex potential landscape, which is achieved by pumping the PCNCs with multiple laser fields. 

The central component of our trap is a photonic crystal nanobeam cavity (PCNC) (Fig. \ref{fig:1}a). PCNCs have already been exploited to realize an efficient near-field trap \cite{Mandal.2010, Lin.2013} as they provide strong optical gradient forces through the evanescent fields. The PCNCs can also exhibit strong SIBA effects because their high-Q resonant optical modes can dynamically respond to a slight change in the local dielectric environment by an external dielectric object. Specifically, a nearby dielectric particle increases the effective optical path length near the cavity, thus decreasing the resonance frequencies of the cavity. This resonance frequency shift becomes more prominent as the particle approaches closer to the PCNC, resulting in a position-dependent frequency shift of the PCNC. It has been previously observed that the displacement of a nearby glass nanoparticle by 100 nm can result in the shift of the PCNC's frequency shift by 1 GHz \cite{Magrini.2018}. Considering typical linewidths of the PCNC’s resonances are on the order of GHz \cite{Deotare.2009},    the shift is enough to change the intensity of the cavity field significantly. 

A prerequisite for quantum optomechanics experiments is levitating the particle without direct contact with the PCNCs. Otherwise, surface friction would immediately destroy any quantum coherence of the particle. However, with a single PCNC, surface contact would be unavoidable, as an optical gradient force from a PCNC is always attractive. We solve this problem by additionally employing another PCNC on the other side of the particle (Fig. \ref{fig:1}b). This scheme first allows for balancing the attractive force from one PCNC with the same pull from the other. Next, suppose the particle is slightly displaced from the force equilibrium position. In that case, the SIBA effects will induce resonance frequency shifts of the two PCNCs in opposite directions, creating a drastic imbalance of the cavity field strengths and respective optical pulling forces. When the detunings of the laser pumps are chosen appropriately, the sum of the two forces can result in a net restoring force that pushes the particle back to its original position (Fig. \ref{fig:1}c). This mechanism makes it possible to stably keep the particle fully levitated from the two PCNCs.

To investigate the feasibility of the PCNC-based SIBA trap, we perform a series of numerical analyses with realistic PCNC designs and experimental parameters. As a device used to create the SIBA trap, we consider a silicon PCNC with a tapered hole array (Fig. \ref{fig:1}a) \cite{Deotare.2009}. Silicon PCNCs have been used extensively in various applications \cite{Eichenfield.2009, Mandal.2010, Liang.2013} and proven to exhibit quality factors up to $1.22\times10^6$ (or intrinsic cavity decay rate down to $\kappa_{in}/2\pi$ = 160MHz) at telecommunication wavelengths \cite{Chan.2012}. We use finite element method (FEM) simulations to design and optimize a free-standing silicon PCNC to have its resonance wavelength at 1550 nm. The simulated intrinsic cavity loss rate is found to be 40 MHz, which is hard to achieve from real devices due to imperfections and contaminations associated with fabrication processes. We instead use the previously demonstrated value of 160 MHz \cite{Chan.2012} as a realistically achievable cavity loss. Considering additional loss channels introduced by nearby PCNC and particle (see \hyperref[sec:supplement]{supplementary information}), the total cavity internal loss rate is assumed to be $\kappa_{in}/2\pi$ = 370 MHz. In addition, we assume the coupling rate between the cavity and the input optical mode ($\kappa_{ex}$) is $2\pi\cdot$80 MHz, resulting in the total cavity loss rate $\kappa/2\pi = \kappa_{in}/2\pi + \kappa_{ex}/2\pi$ = 450 MHz.

\begin{table*}[t]
\caption{List of the parameters and their assumed values for the simulations}
\label{tab:1}
\centering
\begin{tabular}{|l|l|l|}
\hline
\textbf{Parameter}                                                                             & \textbf{Value}                                         & \textbf{Note}                                                                                                                                                             \\ \hline
\begin{tabular}[c]{@{}l@{}}Cavity resonance\\ wavelenght\end{tabular}                          & $\lambda_c \sim$ 1550 nm                               & \begin{tabular}[c]{@{}l@{}}The resonances are assumed \\ to be slightly different for \\ each PCNC\end{tabular}                                                           \\ \hline
Intrinsic cavity loss rate                                                                     & $\kappa_{in}$ = 370 MHz                                &                                                                                                                                                \\ \hline
Total cavity loss rate                                                                         & $\kappa $ = 450 MHz                      &                                                                                                                                                     \\ \hline
\begin{tabular}[c]{@{}l@{}}Distance between the two \\ PCNCs (surface-to-surface)\end{tabular} & d = 600 nm                                             &                                                                                       \\ \hline
Input power                                                                                    & P = 20 uW                                              & For each PCNC                                                                                                                                                                  \\ \hline
Laser detuning                                                                                 & $\delta = \kappa/\sqrt(3) + \Delta\omega_c(\vec{r}_0)$ & \begin{tabular}[c]{@{}l@{}}$\Delta\omega_c(\vec{r}_0)$: the cavity resonance \\ frequency shift by the particle \\ at its origin $\vec{r}_0=(x_0, y_0, z_0)$\end{tabular} \\ \hline
\end{tabular}
\end{table*}

Next, we examine how the PCNC is influenced by a nearby dielectric nanoparticle, i.e., the shift of the PCNC’s cavity frequency as a function of the particle’s position. Here we consider a silica nanosphere with a diameter of 150 nm as the particle, which is widely used in quantum optomechanics experiments based on optical tweezers \cite{Delic.2019, Magrini.2021, Tebbenjohanns.2021}. The cavity frequency shift $\Delta\omega_c(\vec{r})$ can be obtained by conducting the FEM simulation of the PCNC with the particle placed at a given position $\vec{r}$. However, obtaining a full map of $\Delta\omega_c(\vec{r})$ requires repeating the simulations while sweeping the particle position as a parameter, which is computationally costly. Instead, we choose to use the following approximate expression derived from perturbation theory \cite{RomeroIsart.2010}:

\begin{equation}
\label{eq:wshift}
    \Delta\omega_c(\vec{r})\approx-\frac{\omega_0}{2}\frac{(\varepsilon_p-1)|\vec{E}(\vec{r})|^2 V_p}{\int_{PCNC}\varepsilon_m(\vec{r}')|\vec{E}(\vec{r}')|^2 dV' + \int_{air}|\vec{E}(\vec{r}')|^2dV'}
\end{equation}

where $\omega_0$ is the unperturbed cavity frequency, $\varepsilon_p$ and $\varepsilon_m$ are dielectric constants of the particle and the PCNC (i.e., silicon) respectively, $V_p$ is the volume of the particle, and $\vec{E}\left(\vec{r}\right)$ is the electric field of the cavity mode in the absence of the particle. The validity of Eq. \ref{eq:wshift} is confirmed by comparing the values obtained from it with the results from the brute-force simulation including the particle for several particle positions (see \hyperref[sec:supplement]{supplementary information}, Fig. \ref{fig:S1}). Eq. \ref{eq:wshift} allows us to obtain a complete map of $\Delta\omega_c(\vec{r})$ from a single simulation of the cavity field distribution of the unperturbed PCNC, drastically reducing computational overhead. Fig. \ref{fig:2}a shows the two-dimensional plane cut of $\Delta\omega_c(\vec{r})$ across the cavity center. When the particle is brought within 300 nm from the surface of the PCNC, a frequency shift of around 6 GHz is anticipated. Moreover, an additional displacement of the particle’s position by 5 nm is enough to shift the cavity’s frequency more than the assumed linewidth of 450 MHz (Fig. \ref{fig:2}b). Therefore, a strong SIBA effect is expected already at the distance of 300 nm.

As described earlier, our trap consists of two parallel PCNCs placed sufficiently close to each other. We assume the two PCNCs are nearly identical, having the same cavity loss rate and frequency shift response (i.e., $\Delta\omega_c(\vec{r})$) to the particle. However, we require the PCNCs to have different resonance frequencies such that the evanescent fields from the two cavities, when pumped with the same detuning from each resonance, do not result in interference. The surface-to-surface distance of the two PCNCs is assumed to be 600 nm. At this distance, the influence of the two off-resonant PCNCs on each other is found to be negligible, only causing a slight increase in the loss rate (see \hyperref[sec:supplement]{supplementary information}). All the important parameters and conditions used in the analysis are summarized in Table \ref{tab:1}.

The optical force by the PCNC on the particle can be directly computed from the shift of the PCNC’s frequency \cite{Neumeier.2015} and is given by

\begin{subequations}
\label{eq:2}
\begin{equation}
    \vec{F}(\vec{r})=\hbar n (\vec{r}) \vec{\nabla}\Delta\omega_c(\vec{r})
\end{equation}

\begin{equation}
    n(\vec{r})=\frac{P}{\hbar\omega_L}\frac{\kappa_{ex}}{(\kappa/2)^2+(\delta-\Delta\omega_c(\vec{r}))^2}
\end{equation}
\end{subequations}

where n is the intracavity photon number, which depends on the parameters given in Table \ref{tab:1} and the laser frequency $\omega_L$. The above expression is obtained by ignoring the finite response time of the intracavity photon number to the particle-induced frequency shift \cite{Neumeier.2015} and is valid as long as the bandwidth of the cavity ($\kappa/2\pi=$450 MHz) is much faster than the resulting oscillation frequency of the trapped particle, which will be confirmed later.

\begin{figure*}[t]
\centering\includegraphics[width=1.4\columnwidth]{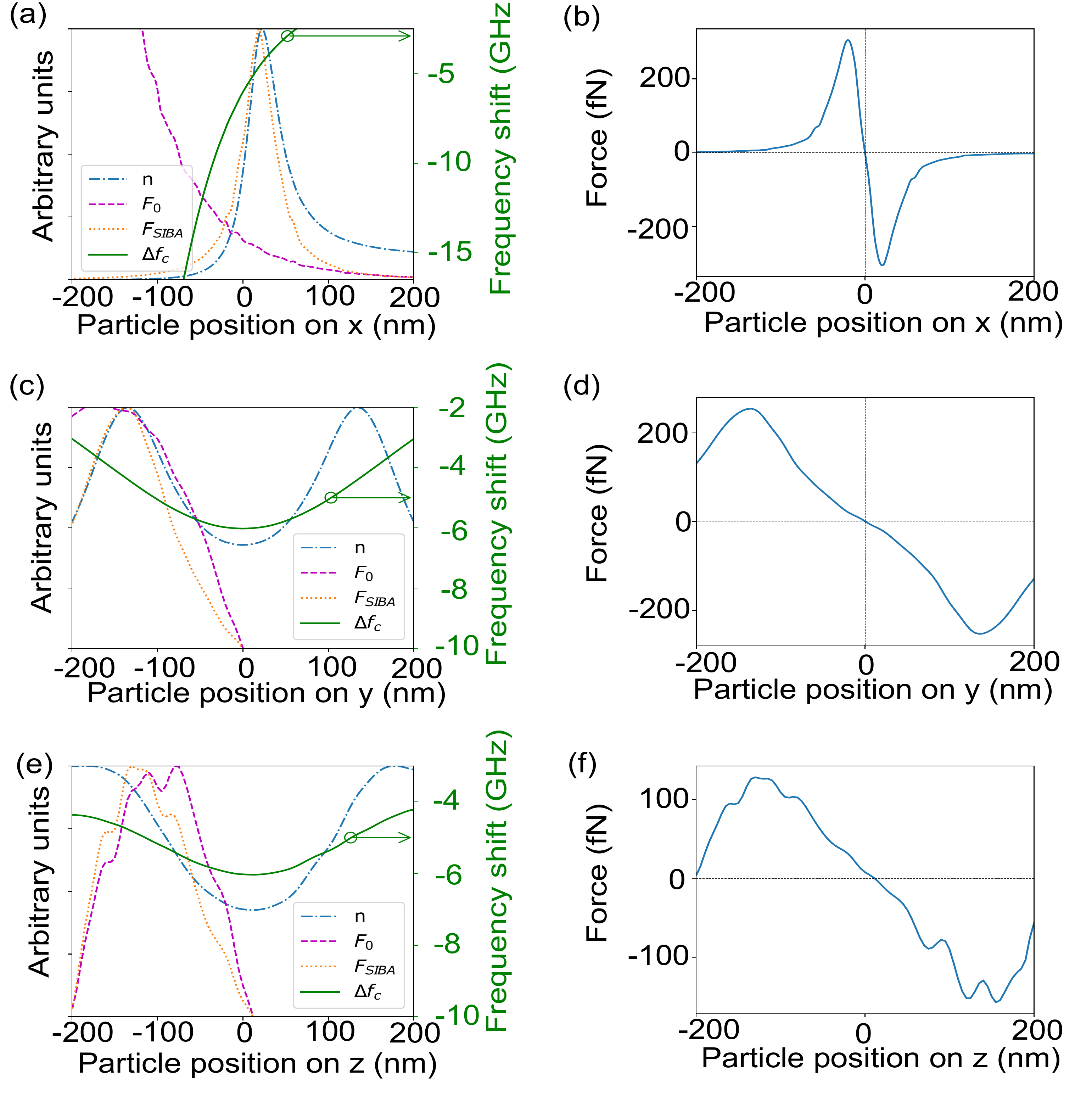}
\caption{\textbf{a,c,e.} Cavity frequency shift ($\Delta f_c=\Delta\omega_c/2\pi$; solid line), intracavity photon number (n; dash-dotted line), and resultant optical SIBA force ($F_{SIBA}$; dotted line) by a single PCNC (positioned at $x = -300$ nm) as a function of the particle’s position along the $x$, $y$, and $z$ directions. For comparison, optical gradient force by a single PCNC with a fixed intracavity photon number ($F_0$; dotted line) is also shown. The profiles are chosen to intersect with ($x_0$, $y_0$,  $z_0$). The change of the frequency shift is significant along the $x$ direction (a). Together with a narrow cavity resonance, it strongly modulates the intracavity photon number, thus resulting in the optical SIBA force sharply peaking at a length scale less than 50 nm. The frequency shift gradients along $y$ and $z$ directions, however, are moderate, inducing much weaker SIBA effects (c, e). The trends of the $F_{SIBA}$ in these directions, therefore, do not show significant deviations from $F_0$. \textbf{b,d,f.} Net forces on the particle along $x$, $y$, and $z$ directions, when the particle is placed in the PCNC-based SIBA trap. The centers of the line cuts coincide with ($x_0$, $y_0$,  $z_0$). In all cases, forces on the negative coordinates are positive and vice versa, confirming the restoring action of the forces toward the origin. A strong SIBA effect along the $x$ direction creates a steep force gradient within the narrow region at the trap center (b), while the forces in the other directions show much more gradual changes over larger length scales (d, f).}
\label{fig:3}
\end{figure*}

We now show how the particle can be levitated in our trap. To that end, we calculate the optical forces exerted on the particle (Fig. \ref{fig:3}). Here we consider pumping each PCNC with the laser input power P = 20 uW and the laser detuning $\delta = \kappa/ \sqrt{3} + \Delta\omega_c(\vec{r}_0)$, where $\Delta\omega_c(\vec{r}_0)$ is the cavity’s resonance frequency shift when the particle is precisely in the middle between the two

\noindent PCNCs (i.e., at position $\vec{r}_0=(x_0,y_0,z_0)$ in our coordinate system). Fig. \ref{fig:3}a shows how the SIBA effect dramatically modifies the optical force. The small displacement of the particle at its original position (x = $x_0$ = 0 in Fig. \ref{fig:3}a) can completely shift the cavity’s resonance frequency out of resonance with the laser pump. This effect converts the monotonously increasing force profile into the one strongly peaked around the origin. Due to the nonlinearities of $n(\vec{r})$ and $\Delta\omega_c(\vec{r})$, the force profiles from the two PCNCs are asymmetric and shifted from each other; the sum of the two forces, therefore, gives rise to a net restoring force (Fig. \ref{fig:3}b). Remarkable is that the change of the force occurs sharply within the length scale of less than 50 nm. This sharp, sub-diffraction feature of the force originates from the PCNC’s narrow optical resonance that responds strongly to the particle’s displacement. We also look at the forces along other spatial directions (Fig. \ref{fig:3}d, f) and confirm that they are also arranged to confine the particle in its origin. We note that the length scales over which the forces change are much larger than the one along the $x$ direction because the SIBA effects are much less significant along those directions due to the modest frequency shift gradients (Fig. \ref{fig:3}c, e).

From Eq. \ref{eq:2}, we find that the curl of the SIBA forces is zero, i.e., $\vec{\nabla}\times\vec{F}(\vec{r})=0$, implying that the work difference is path independent. This allows us to define the potential $U(\vec{r})$ by integrating the work from a fixed reference point in space $\vec{r}_{ref}$:

\begin{equation}
\label{eq:pot}
    U(\vec{r}) = -\int_{\vec{r}_{ref}}^{\vec{r}}\vec{F}\cdot d\vec{r}'
\end{equation}

\begin{figure*}[t]
\centering\includegraphics[width=1.4\columnwidth]{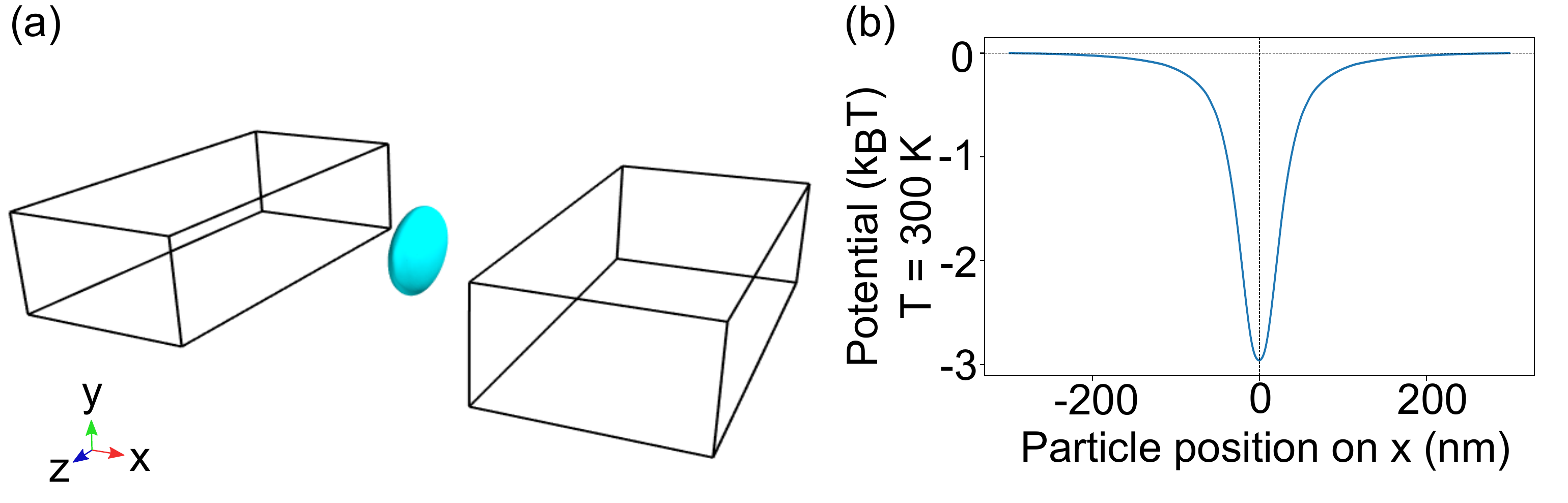}
\caption{\textbf{a.} Three-dimensional potential well formed by the PCNC-based SIBA trap. Shown is the equipotential surface with the potential value of $U=-1k_BT$. The closed surface confirms that the trap can securely contain the particle in three dimensions. The larger extent of the potential along $y$ and $z$ directions is due to modest cavity frequency shift gradients along those directions, which result in weak SIBA effects. \textbf{b.} Potential profile along the $x$ direction. The line cut was taken along $y=y_0$ and $z=z_0$. The width of the potential at half maximum is 56 nm, clearly featuring a length scale much smaller than the diffraction limit. }
\label{fig:4}
\end{figure*}

We numerically compute the potential by choosing the point ($x_{max}$, $y_0$, $z_0$) as the reference point ($\vec{r}_{ref}$), where $x_{max}$=300 nm corresponds to the point closest to the PCNC on the right. Fig. \ref{fig:4} shows that our trap indeed forms a three-dimensional potential well. As already indicated in the force analysis, the full-width-at-half-maximum of the potential along the $x$ direction is 56 nm, an order of magnitude smaller than the diffraction limit (Fig. \ref{fig:4}b). The depth of the potential is sufficiently larger than the thermal energy at room temperature (Fig. \ref{fig:4}b), indicating that the particle can be stably trapped at room temperature. Spring constants of the trap and corresponding particle’s oscillation frequencies can be extracted from the potential and are found to be (15.05 uN/m, 2.14 MHz), (1.15 uN/m, 595.2 kHz), and (0.76 uN/m, 484.12 kHz), respectively, along the $x$, $y$, and $z$ directions. The highest frequency is sufficiently smaller than the cavity decay rate, confirming the validity of the approximation used in Eq. \ref{eq:pot}. 

\begin{figure*}[t]
\centering\includegraphics[width=1.4\columnwidth]{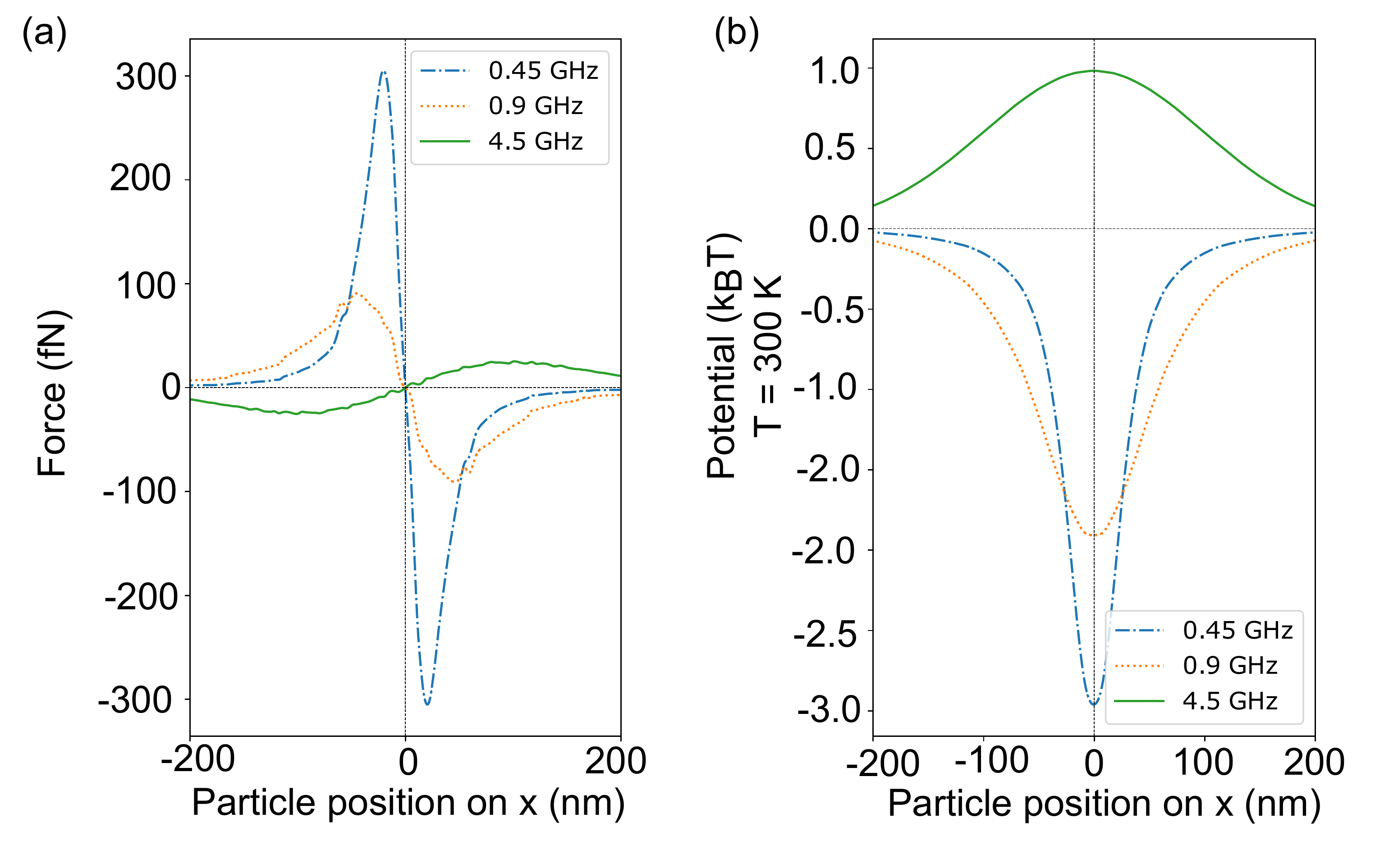}
\caption{\textbf{a.} Trapping forces on the particle along the $x$ direction as a function of the particle’s position for different decay rates $\kappa$: 0.45 GHz, 0.9 GHz, and 4.5 GHz. The line cuts were taken along $y=y_0$ and $z=z_0$. The inset at the left-bottom corner shows the zoom-in of the $\kappa$ = 4.5 GHz case. As the decay rate increases, the magnitude of the restoring force decreases until it disappears. We found the turning point to be around 1.8 GHz. When the decay rate is further increased to 4.5 GHz, the polarity of the force is inverted, and the particle is pushed out from the center. \textbf{b.} The corresponding potentials along the $x$ direction. As the decay rate is increased, the depth of the potential is reduced. When the decay rate is above 1.8 GHz, the potential is flipped such that the center of the potential becomes unstable.}
\label{fig:5}
\end{figure*}

The strong restoring forces in our trap result from the cavity’s sharp resonance, which converts optomechanically induced frequency shift to steep modulation of the cavity field. We observe that the restoring nature of our trap is still maintained when the linewidth is increased from 450 MHz to 900 MHz (Fig. \ref{fig:5}). However, the stiffness and the potential depth are substantially decreased because the cavity with broader resonance does not respond as sharply to the frequency shift. It results in a shallower depth of the trapping potential. When the cavity linewidth exceeds a threshold (1.8 GHz in our design), our trap no longer forms a stable potential and loses the ability to confine a particle.

\begin{figure*}[t]
\centering\includegraphics[width=1.4\columnwidth]{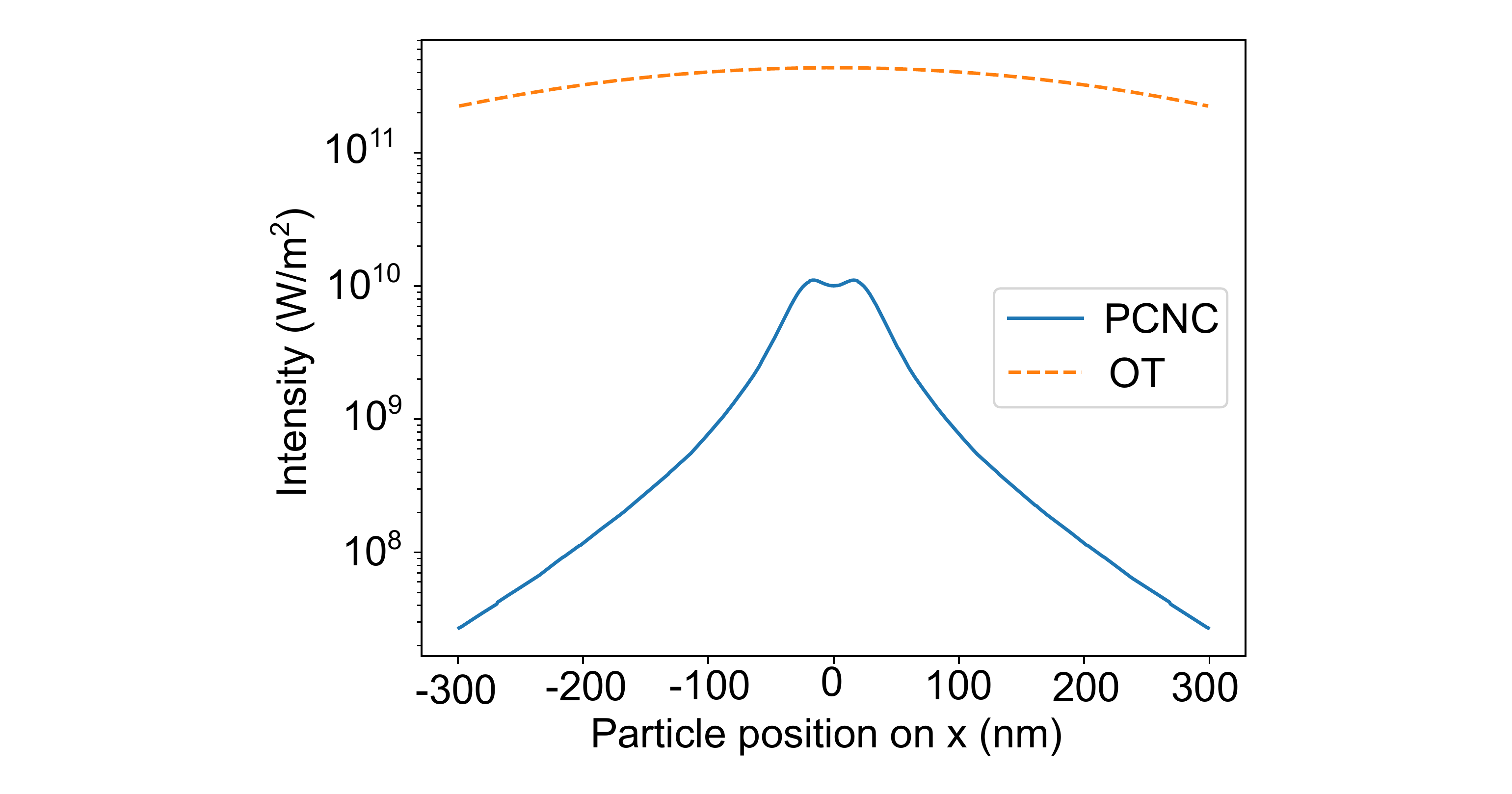}
\caption{Comparison of the optical field intensity that the particle (a silica nanosphere with 150 nm diameter) would see in a standard optical tweezer (OT) and the PCNC-based trap (PCNC). The intensity profiles are calculated assuming that both traps result in the same stiffness (i.e., the spring constant of 15.05 uN/m) along the $x$ direction. The NA of the objective lens for the OT is assumed to be 0.95. The line cuts are taken along $y=y_0$ and $z=z_0$. At the center ($x=0$) of the PCNC-based trap, the particle sees 43 times less intensity compared to the case for the OT. }
\label{fig:7}
\end{figure*}

\begin{figure*}[t]
\centering\includegraphics[width=1.4\columnwidth]{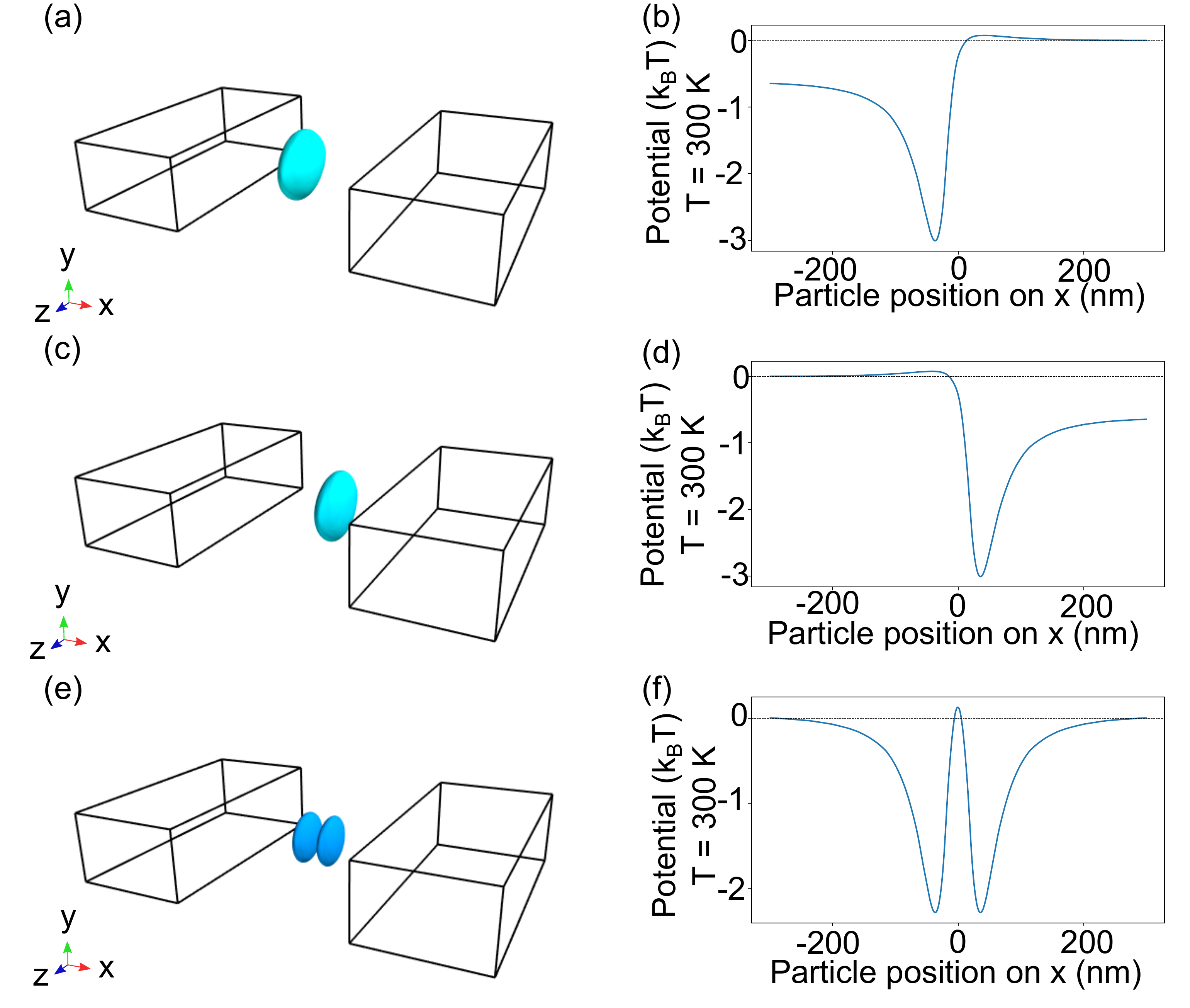}
\caption{\textbf{a.} The equipotential surface of the trap potential formed by choosing the laser detuning for the left PCNC $\delta_L=2\kappa/\sqrt3+\Delta\omega_c(\vec{r}_0)$ and the detuning for the right PCNC $\delta_R=-\kappa/\sqrt3+\Delta\omega_c(\vec{r}_0)$. The value of the potential on the surfaces is fixed to $U=-1k_BT$. \textbf{b.} The line-cut of the corresponding potential along the $x$ direction, taken at $y=y_0$ and $z=z_0$. The position of the trap center is moved to $x = -37$ nm. \textbf{c.} The equipotential surface ($U=-1k_BT$) of the trap potential with $\delta_L=-\kappa/\sqrt3+\Delta\omega_c(\vec{r}_0)$ and $\delta_R=2\kappa/\sqrt3+\Delta\omega_c(\vec{r}_0)$. \textbf{d.} The line-cut of the corresponding potential along the $x$ direction, taken at $y=y_0$ and $z=z_0$. The position of the trap center is $x = 36$ nm. For convenience,  in Figs. c and d, the reference point $r_{ref}$ has been changed to ($x_{min}$, $y_0$, $z_0$), where $x_{min}$=-300 nm. This makes it easier to compare side by side the asymmetric potentials. \textbf{e.} The equipotential surface ($U=-1k_BT$) of the double-well potential formed by driving both PCNCs with two laser tones with the detunings $\delta_1=-\kappa/\sqrt3+\Delta\omega_c(\vec{r}_0)$ and $\delta_2=2\kappa/\sqrt3+\Delta\omega_c(\vec{r}_0)$. \textbf{f.} The line-cut of the corresponding potential along the $x$ direction, taken at $y=y_0$ and $z=z_0$.}
\label{fig:8}
\end{figure*}

Figures \ref{fig:3} and \ref{fig:4} show that our system can form a high-stiffness optical trap (the spring constant of 15.05 uN/m and the corresponding particle’s oscillation frequency of 2.14 MHz) only with 40 uW input power (20 uW input for each PCNC). On the other hand, a standard optical tweezer with a numerical aperture (NA) of 0.95 would need more than three orders of magnitude larger input power (184 mW) to achieve the same trap stiffness (see \hyperref[sec:supplement]{supplementary information}). This drastic reduction in the input power is certainly one of the merits of our trapping scheme. A more important parameter to consider is the optical field intensity that a particle would actually experience in a trap, as it is directly proportional to the absorption heating, which could induce the loss of the particle. We compare the field intensities the particle would see in the case of a standard optical tweezer and our PCNC-based trap (Fig. \ref{fig:7}). Indeed, the field intensity at the center of the PCNC-based trap is a factor of 43 smaller than that of the standard optical tweezer. It primarily results from the difference in the trap width, which is inversely proportional to the square of the spring constant. 

The width of our trap (56 nm) is 18 times narrower than the diffraction-limited width of the standard optical tweezer (995 nm). If our trap were a standard opical trap made with a shorter wavelength laser, the width reduction would result in the reduction of the intensity at the trap center by a factor of $18^2=324$. The reason why this factor is not equal to the observed difference of 43 is that the PCNC-based trap is qualitatively different from standard optical traps. Indeed, Fig. \ref{fig:7} shows another distinctive feature of the PCNC-based trap; in clear contrast to the optical tweezer, the intensity maxima of the PCNC-based trap appear at both sides of the trap center, not at the trap center.

In operating the PCNC-based SIBA trap, the detuning $\delta$ of the input laser field plays a central role as it determines the regions where the cavity field is turned on. We use this principle to obtain further control in trap landscaping. First, we achieve a simple control in displacing the trap by appropriately shifting the input laser detunings to both PCNCs (Fig. \ref{fig:8}a, b, c, d). More fascinating possibilities arise when both cavities are pumped with multiple laser tones with different amplitudes and detunings. One example is a double-well potential shown in Fig. \ref{fig:8}e. We realize it by pumping each PCNC with two laser tones which are used to create two displaced single-well traps. The width of each well and the separation between the wells are 53 nm and 72 nm, respectively, demonstrating the scheme’s ability to create a potential landscape with a nanoscopic resolution.  We anticipate this approach with multiple laser tones will offer a powerful method for creating more complicated potential landscapes beyond the diffraction limit.

% \begin{figure*}[t]
% \centering\includegraphics[width=7cm]{opticafig1}
% \caption{Sample caption (Fig. 2, \cite{Yelin:03}).}
% \end{figure*}

In summary, we have shown that the optical trapping system formed by two parallel silicon PCNCs can levitate a silica nanoparticle with a diameter of 150 nm in a gap between the two PCNCs. Strong optomechanical interactions between the PCNCs and the particle dramatically modify optical force from the cavities, resulting in extremely narrow trapping potential with a width of only 56 nm. The significantly reduced potential width allows for achieving the oscillation frequency of the particle higher than 2 MHz only with a total laser input power of 40 uW. The resulting potential depth, at the same time, is three times higher than the thermal energy at room temperature, indicating that the particle can be trapped stably at room temperature. More remarkable is that the particle in our trap experiences factor of 43 less light intensity than conventional optical tweezers. It ensures that our trap exhibits significantly reduced optical absorption, a crucial requirement for quantum coherent experiments in high vacuum. Finally, by introducing multiple laser tones to each PCNC, we demonstrate the extraordinary versatility of our trapping system in creating complex potential landscapes with nanoscale resolutions. We also note that our PCNC-based trapping system is designed with realistic parameters, therefore, can be realized with available nanofabrication facilities. With all these unique features, our PCNC-based levitation system provides new directions in optical trapping and manipulation, particularly in the field of precision sensing and quantum experiments involving levitated particles.

\textbf{Funding}
The Center for Integrated Quantum Science and Technology (IQST, Johannes-Kepler Grant). The University of Stuttgart.

\textbf{Acknowledgments}
We thank Myungshik Kim and his group members, and Alessio Belenchia for fruitful discussions.

\textbf{Disclosures}
The authors declare no conflicts of interest.

\textbf{Data Availability}
 The data that support this study are available upon reasonable request from the authors.

\textbf{Supplemental document}
See supplemental information for supporting content.

%%%%%%%%%%%%%%%%%%%%%%% References %%%%%%%%%%%%%%%%%%%%%%%%%
\bibliography{main}

\setcounter{figure}{0}
\renewcommand{\thefigure}{S\arabic{figure}}
\setcounter{equation}{0}
\renewcommand{\theequation}{S\arabic{equation}}

\clearpage

%%%%%%%%%%%%%%%%%%%%%%%%%%%%%%%%%%%%%%%%%%%%%%%%%%%%%%%%%%%%%%%%%%%%%%%%%%%%%%%%%%% Supplementary Information %%%%%%%%%%%%%%%%%%%%%%%%%%%%%%%%%%%%%%%%%%%%%%%%%%%%%%%%%%%%%%%%%%%%%%%%%%%%%%%%%%%
\section{Supplementary Information} \label{sec:supplement}

\subsection{Calculation of intensity and power for optical tweezers}

Here we present the model for an optical tweezer that we assumed to obtain the intensity distribution of the optical tweezer for a given particle’s oscillation frequency in the tweezer and the tweezer’s numerical aperture.

We first assume that we are in the Rayleigh scattering regime, which is a valid approximation as the particle size (150 nm) is sufficiently smaller than the laser beam’s wavelength (1550 nm). In this regime, the gradient force acting on the particle is given by \cite{Harada.1996}

\begin{equation}
\label{eq:S1}
    \vec{F} = \frac{2\pi a^3}{c}\left(\frac{m^2-1}{m^2+2}\right)\nabla I(\vec{r})
\end{equation}

where $I$ is the electric field intensity acting on the particle, $a$ is its radius, $c$ is the speed of light, and $m=n_0/n_1$ is the ratio of the refractive indices of the particle and the medium, respectively.

This expression can be simplified by assuming that the surrounding medium is a vacuum, so $n_1=1$ and $m=n_0$. If we further assume that we are dealing with a non-magnetic material, the relation $m^2=\varepsilon_r$ holds, where $\varepsilon_r$ is the relative permittivity of the particle. We also consider that the intensity of the laser beam with waist $W$ has a Gaussian distribution corresponding to the TEM$_{00}$ mode and approximate it for small movements $x$ around the center ($x\ll W$), where the intensity is $I_0$, as \cite{Saleh.1991}

\begin{equation}
\label{eq:S2}
    I(x) = I_0 exp(-2x^2/W^2) \approx I_0(1-2x^2/W^2)
\end{equation}

Inserting Eq. \ref{eq:S2} into Eq. \ref{eq:S1}, we obtain the force along the $x$ direction at around the trap center

\begin{equation}
\label{eq:S3}
    F_x = - \frac{8\pi a^3}{c}\left(\frac{\varepsilon_r-1}{\varepsilon_r+2}\right)I_0x/W^2
\end{equation}

The stiffness $k$ relates to the force $F$ and position $x$ as $F_x=-kx$ and, at the same time, determines the oscillation frequency or trapping frequency $\omega_T$ for an oscillating mass $m$ via $k=\omega_T^2m$. We will also use the volume of the particle approximated as a sphere $V=\frac{4}{3}\pi a^3$, where $a$ is the radius of the particle, and its density $\rho=m/V$. Eq. \ref{eq:S3}, therefore, is converted to

\begin{subequations}
    \begin{equation}
    \label{eq:S4a}
        \omega_T^2 m = \frac{6V}{c}\left(\frac{\varepsilon_r-1}{\varepsilon_r+2}\right)I_0/W^2
    \end{equation}
    
    \begin{equation}
    \label{eq:S4b}
        I_0=\left(\frac{\varepsilon_r+2}{\varepsilon-1}\right)\frac{\rho c \omega_T^2W^2}{6}
    \end{equation}
\end{subequations}

The optical tweezers intensity shown in Fig. 6 of the main text is obtained following Eq. \ref{eq:S2} and Eq. \ref{eq:S4b} derived here, assuming the numerical aperture $NA$ = 0.95, the wavelength of the tweezer $\lambda$=1550 nm, and the geometric relation $W=\lambda/(\pi\cdot NA)$. For a silica particle we use the values $n_0$=1.47, $\varepsilon_r$=2.15, and $\rho$=1850 kg/m$^3$ for the refractive index, dielectric constant, and density, respectively.
To calculate the required power to trap the particle with that laser intensity $I_0$, we used the beam power relation \cite{Saleh.1991}

\begin{equation}
\label{eq:S5}
    P=\frac{1}{2}I_0(\pi W^2)
\end{equation}

\subsection{Comparison between a brute-force approach and an approximation for frequency shift calculations}

Here we calculate the PCNC's resonance frequency shifts by the silica nanosphere (150 nm in diameter) using two different approaches and compare the results.

In the first approach, which we call the brute-force approach, we run the FEM simulation with the nanoparticle included in the model to obtain the resonance frequency in the presence of the particle. Then we take the difference between the result and the resonance frequency obtained without the particle. While this approach is straightforward, it requires repeating the simulations for different positions and sizes of the particle. Therefore, it is unsuitable for obtaining the frequency shift map as a function of the particle's position with high resolution.

\begin{figure}[t]
\centering\includegraphics[width=0.7\columnwidth]{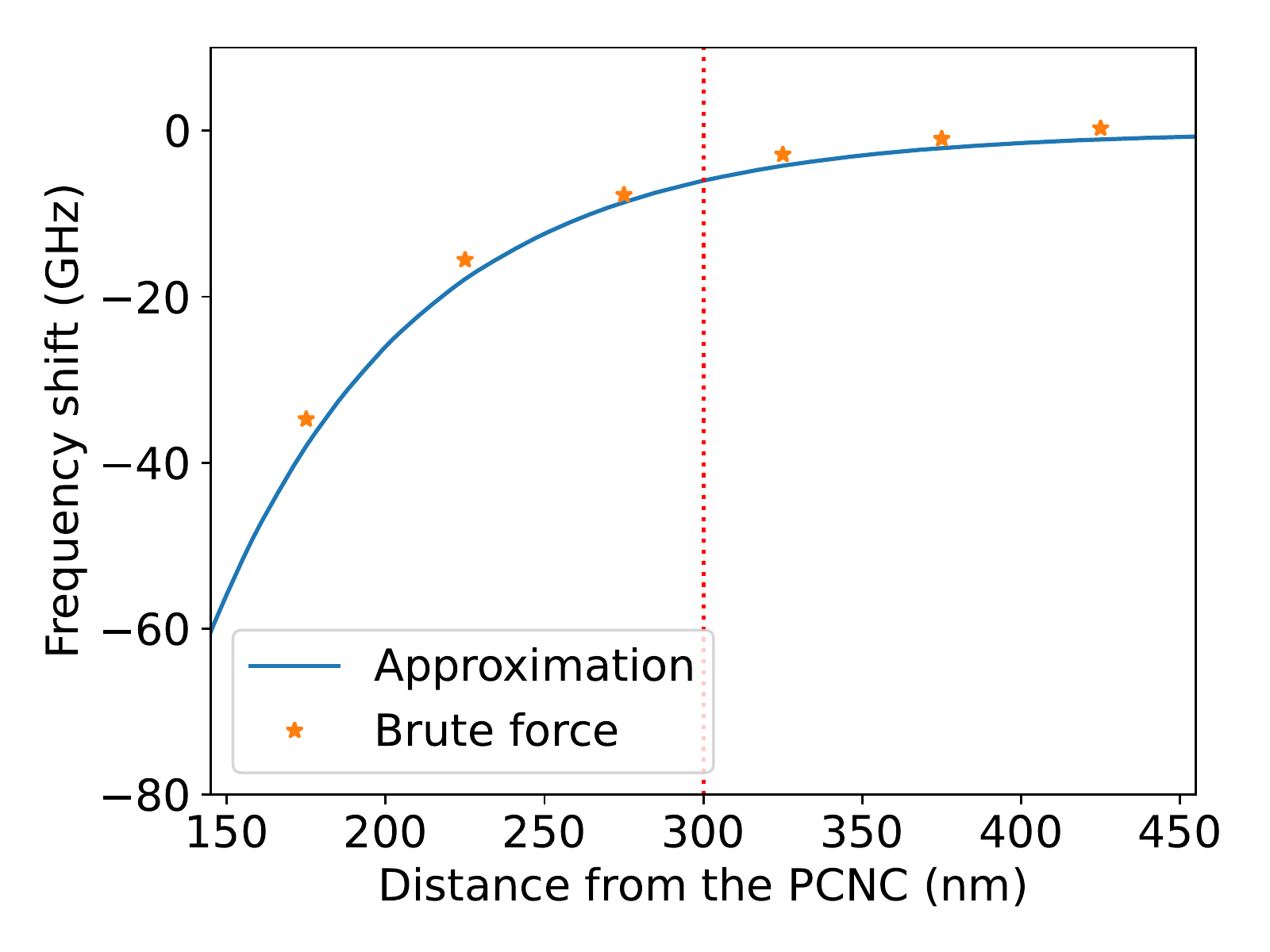}
\caption{The PCNC's resonance frequency shift is calculated by the two different approaches as a function of the particle's distance from the PCNC. The solid blue curve (Approximation) is obtained by using Eq. 1 in the main text. The other approach (*) uses a brute-force FEM simulation that directly takes account of the particle as a part of the domain. Each data point in this approach, therefore, is obtained by performing a new simulation with the updated position of the particle. The vertical red line indicates the nominal distance between the particle and the PCNC (300 nm) when the trap is formed.}
\label{fig:S1}
\end{figure}

\begin{figure*}[t]
\centering\includegraphics[width=1.4\columnwidth]{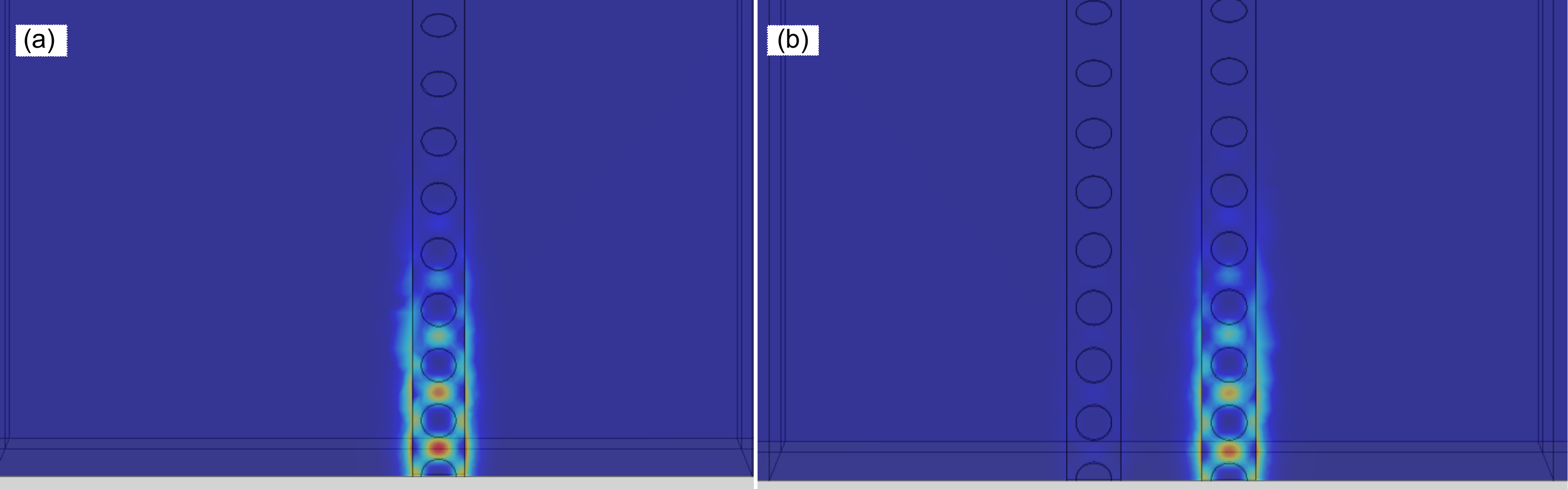}
\caption{FEM simulation of a single PCNC (a), and two PCNCs separated by a distance of 600 nm (b). The mode profile remains the same and we observe the loss rate increase of only $\Delta\kappa/2\pi = 30$ MHz.}
\label{fig:S2}
\end{figure*}

For the second approach, we use an approximation from Romero-Isart et al. \cite{RomeroIsart.2010}. It estimates the cavity’s frequency shift by the particle from the original cavity field in the absence of the particle. We further assume that the electric fields inside the particle’s volume are almost constant and approximated as the value at the particle’s center-of-mass position. The resulting expression is given in Eq. 1 in the main text. The expression allows us to obtain the map of frequency shift as a function of the particle’s position with a single simulation run.

Both approaches are compared in Fig. \ref{fig:S1} along the line defined by the intersection of planes $y_0$ and $z_0$. The results from the two methods are in good agreement, confirming the validity of Eq. 1 we use throughout our study. 

\subsection{Effect of a nanoparticle and the neighboring PCNC on the internal loss rate of the PCNC}

In the main text, we first assume that the intrinsic loss rate of our PCNC is 160 MHz, based on a previously achieved value \cite{Chan.2012}. However, the loss rate increases further if additional structures or materials are added in proximity, perturbing the local dielectric environment. In our case, the nanoparticle and the additional PCNC placed in parallel can contribute to the loss rate increase. Here we assume that the loss channels caused by the particle and the PCNC are independent. Then, we can express the total internal loss rate $\kappa_{in}$ as

\begin{equation}
    \kappa_{in} = \kappa_{in,0} + \kappa_{PCNC} + \kappa_{part}
\end{equation}

where $\kappa_{in,0}$ is the intrinsic loss rate of the bare PCNC, and $\kappa_{PCNC}$ and $\kappa_{part}$ are the additional loss rates caused by the second PCNC and the nanoparticle, respectively. In the main text, we assume $\kappa_{in,0}/2\pi = 160$ MHz. Below, we provide our estimates for $\kappa_{PCNC}$ and $\kappa_{part}$.

To calculate $\kappa_{PCNC}$, we perform two FEM simulations, one with only a single PCNC and the other with two PCNCs separated by 600 nm (as is the condition for the PCNC-based trap). In the latter case, to prevent the resonant coupling between the two PCNCs, the left PCNC is reduced in scale by 0.5\% along the direction of the PCNC. Fig. \ref{fig:S2} shows the mode profiles for the same PCNC in both FEM simulations and confirms that the mode profile remains the same. The simulation with two PCNCs observes an increase in the loss rate of $\Delta\kappa/2\pi=30$ MHz. We, therefore, conclude that $\kappa_{PCNC}/2\pi = \Delta\kappa/2\pi = 30$ MHz.

\begin{figure}[t]
\centering\includegraphics[width=0.7\columnwidth]{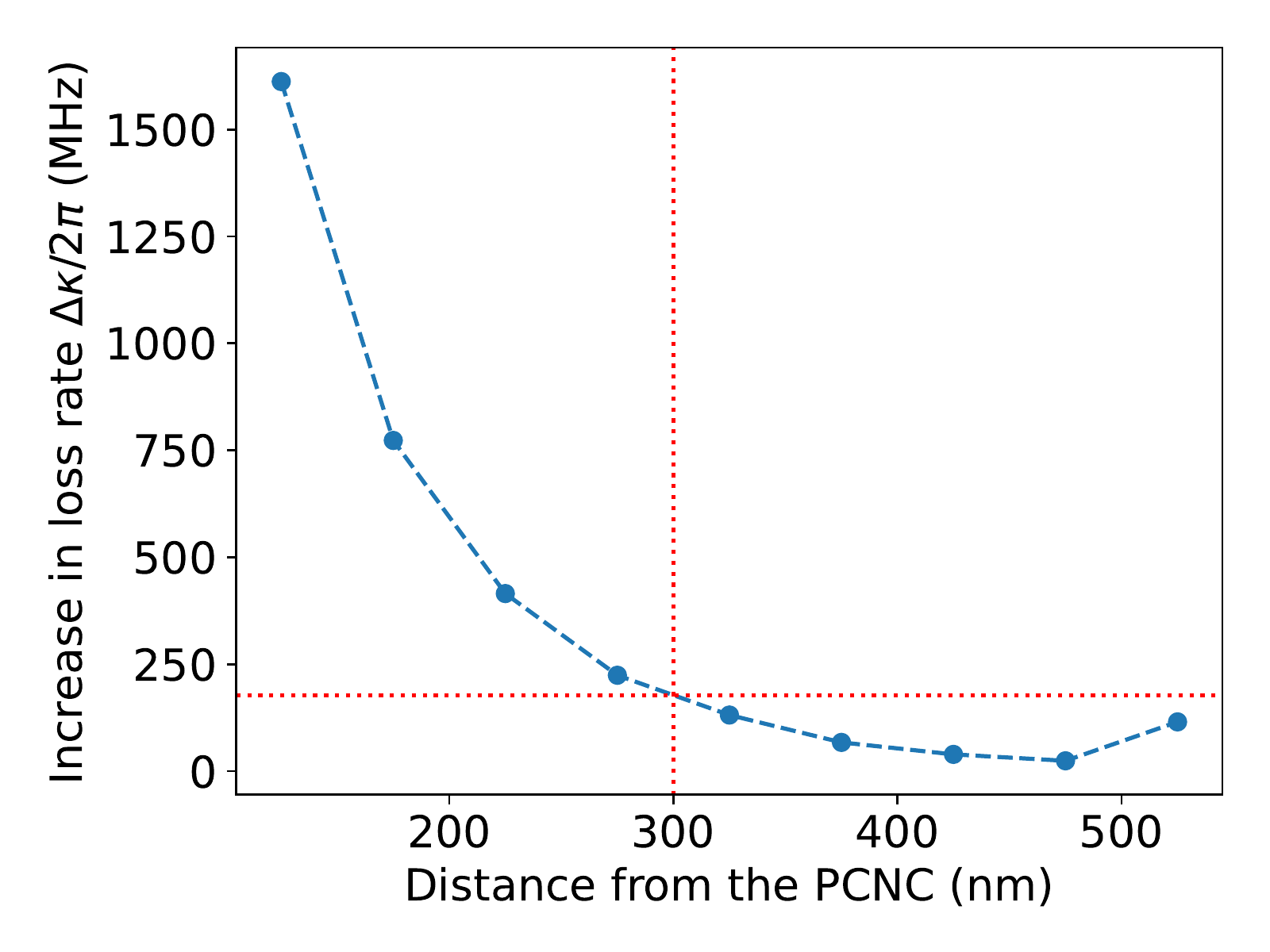}
\caption{Increase in the loss rate $\Delta\kappa/2\pi$ caused by a 150 nm-in-diameter silica nanosphere on the PCNC as a function of the distance between them. The equilibrium position of the nanoparticle in the double-PCNC trapping scheme lies 300 nm away from the PCNC, where $\Delta\kappa/2\pi=$180 MHz.}
\label{fig:S3}
\end{figure}

To calculate $\kappa_{part}$, we perform the same brute-force FEM simulation used in Fig. \ref{fig:S1}. Instead of focusing on the frequency shift, we compute the increase in the loss rate compared to the original loss rate obtained without the particle. Fig. \ref{fig:S3} shows how the loss rate increases as the particle moves close to the PCNC. Around the equilibrium position of the nanoparticle in the trap (300 nm away), the measured increase in loss rate is $\Delta\kappa/2\pi=180$ MHz. We choose this value to be the nominal value for $\kappa_{part}$.

In summary, the total intrinsic loss rate is $\kappa_{in}/2\pi = 160$ MHz + 30 MHz + 180 MHz = 370 MHz.

\end{document}